\documentclass[pre,twocolumn]{revtex4}
\usepackage{amsmath,longtable}
\usepackage{rotating}
\usepackage{graphicx}

\begin{document}

\newcommand{\suppress}[1]{}
\renewcommand{\marginpar}[1]{}
\newcommand{\sign}{\mbox{sign}}
\newcommand{\citen}[1]{\citet{#1}}
\newcommand{\citea}[1]{\citep{#1}}

\title{Predicting Genetic Regulatory Response Using Classification}

\author{Manuel Middendorf\,$^{\rm a}$, Anshul Kundaje\,$^{\rm b}$,\\ Chris Wiggins\,$^{\rm c,d}$, Yoav Freund\,$^{\rm b,d,e}$,\\Christina Leslie\,$^{\rm b,d,e}$\\
{\footnotesize \em $^{\rm a}$Department of Physics, $^{\rm b}$Department of Computer Science,\\ $^{\rm c}$Department of Applied Mathematics,\\ $^{\rm d}$Center for Computational Biology and Bioinformatics,\\ $^{\rm e}$Center for Computational Learning Systems,\\ Columbia University, NY 10027, USA}
}

\begin{abstract}

\noindent {\bf Motivation:}
Studying gene regulatory
mechanisms in simple model organisms
through analysis of high-throughput genomic data
has emerged as a central
problem in computational biology.  Most approaches in the literature
have focused either on finding a few strong regulatory patterns
or on learning descriptive models from training data.
However, these approaches are not yet adequate for making
accurate predictions about which genes will be up- or down-regulated
in new or held-out experiments.  By introducing a predictive methodology
for this problem, we can use powerful tools from machine learning
and assess the statistical significance of our predictions.
\\
{\bf Results:}
We present a novel classification-based method for learning to predict
gene regulatory response.
Our approach is motivated by the hypothesis that
in simple
organisms such as {\it Saccharomyces cerevisiae}, we can learn
a decision rule for predicting whether a gene is up- or
down-regulated in a particular experiment based on (1)
the presence of binding site
subsequences
(``motifs'')
in the gene's regulatory region
and (2) the expression levels of regulators such as
transcription factors in the experiment (``parents'').
Thus our learning task
integrates
two
qualitatively different data sources:
genome-wide cDNA microarray data
across multiple perturbation and mutant experiments along with
motif profile data from regulatory sequences.
We
convert
the regression task of
predicting real-valued gene expression measurements
to a
classification task of predicting +1 and -1 labels, corresponding to
up- and down-regulation beyond the levels of biological and
measurement noise in microarray measurements.
The learning algorithm employed is boosting with a margin-based generalization of decision trees, alternating decision trees. This large-margin classifier is sufficiently flexible to allow complex logical functions, yet sufficiently simple to give insight into the
combinatorial mechanisms of gene regulation.
We observe
encouraging
prediction accuracy on experiments based on the Gasch
{\it S. cerevisiae} dataset, and we show that we can
accurately predict up- and down-regulation on held-out experiments.
We also show how to extract significant regulators, motifs, 
and motif-regulator pairs from the learned models for various
stress responses.
Our method thus provides predictive hypotheses, suggests
biological experiments, and provides interpretable insight into
the
structure
of genetic regulatory networks.
\\
{\bf Availability:} The MLJava package is available by
request from the authors.\\
{\bf Contact:} cleslie@cs.columbia.edu\\
{\bf Supplementary data:} http://www.cs.columbia.edu/compbio/geneclass

\end{abstract}

\maketitle

\section{Introduction}
Understanding underlying mechanisms of gene transcriptional regulation through
analysis of high-throughput genomic data
-- for example,
gene expression
data from microarray experiments and regulatory sequence data --
has
become one of the central problems in computational biology, particularly
for simpler model organisms such as {\em S. cerevisiae}.
Efforts to identify regulatory elements in non-coding DNA \citea{bussemaker:reduce,hughes:alignace},
models for investigating co-occurrence of regulatory motifs and combinatorial effects of regulatory molecules \citea{pilpel:motifsyn}, and attempts to organize genes that appear to be subject to common regulatory control into ``regulatory
modules'' \citea{ihmels:module,segal:module} all
define
pieces of this complex problem.
Most recent
studies of
transcriptional regulation can be placed broadly
in one of three categories: {\em statistical approaches}, which aim to
identify
statistically significant regulatory
patterns in a dataset
\citea{bussemaker:reduce, pilpel:motifsyn, ihmels:module}; {\em probabilistic approaches},
which
try to discover structure in the dataset as formalized by probabilistic models
(often graphical models or Bayesian networks) \citea{segal:module, segal:learningmod, hartemink:graphical, peer:inferring, peer:minreg}; and
{\em linear network models}, which hope to learn explicit parameterized
models for pieces of the regulatory network by fitting to data \citea{collins:svd,hassle:lin}.  These approaches are all useful exploratory tools
in the sense that they
allow the user to generate
biological hypotheses about transcriptional regulation that can then be
tested in
the lab.
In general, however, these approaches
are not yet adequate
for making accurate {\em predictions} about which genes will be up- or
down-regulated in new or held-out experiments.
Therefore, it is
difficult to compare performance of different approaches or decide, based
on cross-validation experiments, which approach is likely to generate
plausible hypotheses.

The goal of our method is to learn a prediction function for the
regulatory response of genes under different experimental conditions.
The inputs to our learning algorithm are
the gene-specific regulatory sequences -- represented by the set of
 binding site patterns they contain (``motifs'') -- and
the experiment-specific expression levels of regulators (``parents'').
The output
is a prediction of the
expression state 
of
the regulated gene.
Rather than trying to predict a real-valued expression level,
we formulate the task as a binary classification problem, that is,
we predict only whether the gene is up- or down-regulated.
This reduction allows us to exploit modern and effective
classification algorithms.
The learning algorithm that we use is boosting with
a margin-based generalization of decision trees called
alternating decision trees (ADTs).  Boosting, like support vector
machines, is a large-margin classification algorithm that performs
well for high-dimensional problems.
We evaluate the performance of our method
by measuring prediction accuracy on
held-out microarray experiments, and we achieve very good classification
results in this setting.
Moreover, we show that the learned prediction trees contain information
that is
both
statistically significant and biologically meaningful.
These significant
features, which are associated with accurate generalization rather
than simply correlations in the training data,
suggest regulators, motifs, and motif-regulator pairs 
that
play an important role in gene transcriptional regulation.

Among recent statistical approaches, the most
successful related approach
is the REDUCE algorithm of \citen{bussemaker:reduce} for regulatory
element discovery.  Given
gene expression measurements from a single microarray experiment and the
regulatory sequence $S_g$ for each gene $g$ represented on the array,
REDUCE proposes a linear model for the dependence of log gene expression
ratio $E_g$
on presence of regulatory
subsequences
(or ``motifs")
\newcommand{\M}{\mu}
$E_g = C + \sum_{\M \in S_g} F_\M N_{\M g}$, where $N_{\M g}$ is a count of occurrences of regulatory
subsequence
$\M$ in sequence $S_g$, and the $F_\M$ are experiment-specific fit parameters.
We generalize beyond the conditions of a single experiment
by considering pairs (motif$_g$,parent$_e$), where the parent variable
represents over- or under-expression of a regulator (transcription factor, signaling molecule, or protein kinase) in the experiment $e$, rather than
using motif information alone.
Note, however, that we use classification rather than regression as in REDUCE.

Similar restriction of potential parents
has been used with success
in the probabilistic model literature, including
in the regression-based work of
\citen{segal:module}
for partitioning target genes into {\em regulatory modules} for
{\em S. cerevisiae}.  Here, each module is
a probabilistic
regression tree, where internal nodes of
the tree correspond to states of regulators
and each
leaf node prescribes a normal distribution
describing the expression of all the module's genes
given the regulator conditions.  The authors provide
some statistical validation on new experiments by
establishing that selected module distributions do have non-random
correlation
with true expression;
however,
they do not focus on making accurate predictions of differential
expression as we
do here.
In our work, we retain the distinction between regulator (``parent'') genes
and target (``child'') genes, as well as a
model that can
capture combinatorial relationships among regulators;
however, our margin-based trees are very different from
probabilistic trees. Unlike
in \citen{segal:module}, we learn from both expression and sequence data, so that the influence of a regulator is mediated through presence of regulatory
element.  We note that in separate work,
\citen{segal:learningmod} present a
probabilistic model for combining
promoter sequence
data and a large amount of expression data to learn transcriptional
modules on a genome-wide level in {\em S.\ cerevisiae},
but they do not demonstrate how to use this learned model
for predictions of regulatory response.

\suppress{
}

\section{Learning algorithm}
\label{sec:learning}
\subsection{Adaboost}
\label{sec:adaboost}

Adaboost is a general discriminative learning algorithm
invented by Freund and
Schapire~
\citea{Schapire02}.
The basic idea of Adaboost is
to repeatedly apply a
simple learning algorithm, called the {\em weak} or {\em base} learner,
to different weightings of the same training set. In its simplest form,
Adaboost is intended for binary prediction problems
where the training set consists of pairs
$(x_1,y_1),(x_2,y_2),\ldots,(x_m,y_m)$, $x_i$ corresponds to the
features of an example, and $y_i \in \{-1,+1\}$ is the binary label
to be predicted. A {\em weighting} of the training examples is an
assignment of a non-negative real value $w_i$ to each example
$(x_i,y_i)$.

On iteration $t$ of the boosting process, the weak learner is  applied
to the training set with a set of weights $w_1^t,\ldots,w_m^t$ and
produces a prediction rule $h_t$ that maps $x$ to
$\{0,1\}$.
The requirement on the weak learner is for
$h_t(x)$ to have a small but significant correlation
with the example labels $y$  when measured using the
{\em current weighting of the examples}. After the rule $h_t$ is
generated, the example weights are changed so that the weak
predictions $h_t(x)$ and the labels $y$ are decorrelated. The weak
learner is then called with the new weights over the training examples
and the process repeats.  Finally, all of the weak prediction rules
are combined into a single {\em strong} rule using a weighted majority
vote.  One can prove that if the rules generated in the iterations are
all slightly correlated
with the label, then the
strong rule will have a very high correlation with the label -- in other
words, it will predict the label very accurately.

The whole process can be seen as a variational method in which an
approximation $F(x)$ is repeatedly changed by adding to it small
corrections given by the weak prediction functions.
In
Figure~\ref{fig:Adaboost}, we describe Adaboost in these terms. We
shall refer to $F(x)$ as the {\em prediction score} in the rest of the
paper. The strong prediction rule learned by Adaboost is
$\mbox{sign}(F(x))$.

A surprising phenomenon associated with Adaboost is that the test
error of the strong rule (percentage of mistakes made on new examples)
often continues to decrease even after the training error (fraction of
mistakes made on the training set) reaches zero.
This behavior has been related to the concept of a
``margin'', which is simply the value $y F(x)$~\citea{SchapireFrBaLe98}.
While $y F(x)>0$ corresponds to a correct prediction, $y F(x)>a>0$
corresponds to a {\em confident} correct prediction, and the
confidence increases monotonically with $a$.
Our experiments in this
paper demonstrate the correlation between large margins and correct
predictions on the test set
(see Results section).

\begin{figure}[t]
\begin{center}
\begin{tabular}{|l|}
\hline
 $F_0(x)\equiv0$\\
 for $t=1\ldots T$\\
\hspace{1 cm} $w_i^t=\exp(-y_iF_{t-1}(x_i))$\\
\hspace{1 cm} Get $h_t$ from {\it weak learner}\\
\hspace{1 cm} $
\alpha_t=\ln\left(\frac
        {\sum_{i:h_t(x_i)=1, y_i=1} w_i^t}
        {\sum_{i:h_t(x_i)=1, y_i=-1} w_i^t}
\right) $\\
\hspace{1 cm} $F_{t+1}=F_t+\alpha_th_t$\\
\hline
\end{tabular}
\end{center}
\caption{{\footnotesize {\bf The Adaboost algorithm.}} \label{fig:Adaboost}}
\end{figure}

\subsection{Alternating Decision Trees}
\label{sec:adt}

\begin{figure}[htb]
\begin{center}
\includegraphics[scale=0.32]{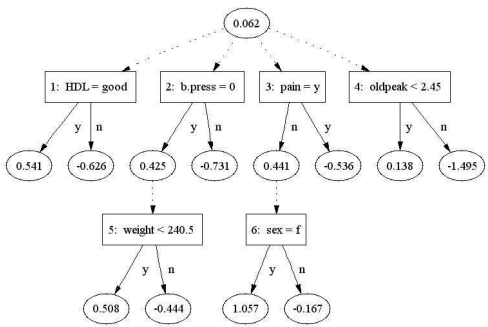}
\caption{\footnotesize {\bf An example ADT.}}\label{fig:adtexample}
\end{center}
\end{figure}

Adaboost is often used with a decision tree learning algorithms as the
base learning algorithm.~
We use Adaboost both to learn the decision rules constituting the tree
and to combine these rules through a weighted majority vote. The form
of the generated decision rules is called an {\em alternating decision
tree} (ADT)~\citea{ freund:alt}.

We explain the structure of ADTs using the example given in
Figure~\ref{fig:adtexample},  reproduced from~\citen{freund:alt}.
The problem domain is heart disease diagnostics and the goal is to
predict whether an individual is healthy or sick based on 13 different
indicators.  The tree consists of alternating levels of ovals ({\em
prediction nodes}) and rectangles ({\em splitter nodes}).
The numbers within the ovals
define contributions to the prediction score. In this example,
positive contributions are evidence of a healthy heart, negative
contributions are evidence of a heart problem.  To evaluate the
prediction for a particular individual we start at the top oval
($0.062$) and follow the arrows down. We follow {\em all} of the
dotted arrows that emanate from prediction nodes, but we follow
{\em only one}
of the solid-line arrows emanating from a splitter node,
corresponding to the answer (yes or no) to the condition stated
in rectangle. We sum the values in all the prediction nodes
that we reach. This sum represents the prediction score $F(x)$ above,
and its sign is the prediction.

For example, suppose we had an individual for which
{\sc hdl=bad},
{\sc b.press=0}, {\sc pain=y},
{\sc oldpeak=2}, {\sc weight=300}, {\sc sex=m}.
In this case, the prediction
nodes that we reach in the tree are
$0.062$, $-0.626$, $0.425$, $-0.444$, $-0.536$, $0.138$, and summing 
gives a
score of $-0.981$, i.e., a very
confident diagnosis that the individual has a heart problem.

The ADT in the figure was generated by Adaboost from training data.
In terms of Adaboost,
each prediction node represents a weak prediction rule, and at
every boosting iteration, a new splitter node together with its two
prediction nodes is introduced. The splitter node can be attached to
any previous prediction node, not only leaf nodes.  Each prediction
node is associated with a weight $\alpha$ that contributes to the
prediction score of every example reaching it.  The weak hypothesis
$h(x)$ is 1 for every example reaching the prediction node and 0 for
all others.
The number in front of the conditions in the splitter nodes of
Figure~\ref{fig:adtexample} indicates the iteration number on which
the node was added. In general, lower iteration numbers indicate that
the decision rule is more important.  We use this heuristic to
analyze the ADTs and identify the
most important factors in gene regulatory response.

\subsection{ADTs for Predicting Regulatory Response}
\label{sec:adtforpred}

\begin{figure}[htb]
\begin{center}
\resizebox{!}{2.5in}{\includegraphics{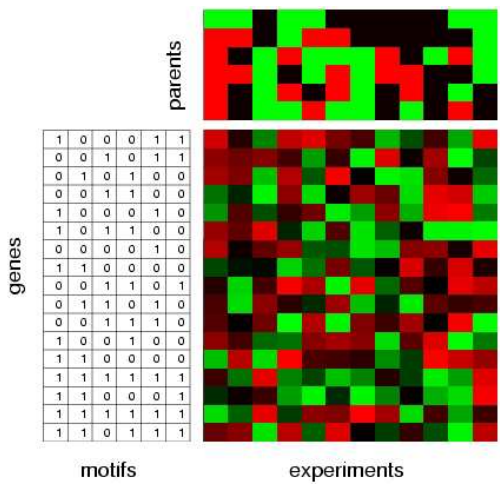}}
\end{center}
\caption{\footnotesize {\bf Representation of data for regulatory response prediction.}  Every (target gene,experiment) is assigned a label of
+1 (up-regulated, in red) or -1 (down-regulated, in green) and
represented by the gene's vector of
motif counts (only binary values shown here) and the experiment's vector of regulator expression states.
\label{fig:datarep}}
\end{figure}

\renewcommand{\P}{\pi}
For the problem of predicting differential gene expression, we start
with a candidate set of {\em motifs} $\M$ representing known or putative
regulatory element sequence patterns and a candidate set of regulators or
{\em parents} $\P$.
For each (gene,experiment) example in our gene expression dataset,
we have two sources of feature information relative to the candidate
motifs and candidate parent sets: a vector $N_{\M g}$ of motif counts of occurrences
of patterns $\M$ in the regulatory sequence of gene $g$,
and the vector $\P_e\in\{-1,0,1\}$
of expression states for parent genes $\P$ in the experiment $e$.
The data
representation is depicted in Figure \ref{fig:datarep}.

Splitter
nodes in our ADTs depend on decisions based on (motif,parent) pairs.
However, instead of splitting on real-valued thresholds of parent expression
and integer-valued motif count thresholds, we consider only whether a
motif $\M$ is present or not, and only whether a parent $\P$ is
over-expressed (or under-expressed) in the example.  Thus, splitter
nodes make boolean decisions based on conditions such as ``motif
$\M$ is
present and regulator $\P$ is over-expressed
(or under-expressed)''.
Paths in the learned ADT correspond to
conjunctions (AND operations) of these boolean (motif,parent) conditions.
Full details on selection of the candidate
motifs and regulators and discretization into up and down states is given
in the Methods section.

\section{Methods}
\label{sec:methods}

\indent{\bf\em Dataset:} We use the \citen{gasch:yeast}
environmental stress response dataset,
consisting of
cDNA microarray experiments measuring genomic expression in
{\em S. cerevisiae} in response to diverse environmental
transitions.
There are a total of
6110 genes and 173 experiments
in the dataset,
with
all measurements given as $\log_2$ expression values (fold-change with
respect to
unstimulated reference expression).
We do not perform a zero mean
and unit variance normalization over experiments,
since we must retain the meaning of the true zero (no fold change).

{\bf\em Motif set:} We obtain the 500 bp 5' promoter
sequences of all \emph{S. cerevisiae} genes from the Saccharomyces Genome
Database (SGD). For each of
these sequences, we search for transcription factor (TF) binding
sites using the PATCH software licensed by TRANSFAC
\citea{wigender:transfac}. The PATCH tool uses a library of known
and putative TF binding sites, some of which are represented by
position specific scoring matrices and some by consensus patterns,
from the TRANSFAC Professional database.
A total of 354 binding
sites
are used after pruning to remove redundant and rare sites.

{\bf\em Parent set:} We compile different sets of candidate
regulators to study the performance and dependence of our method
on the set of regulators. We restrict ourselves to a superset of
475 regulators (consisting of transcription factors, signaling
molecules and protein kinases), 466 of which are used in
\citen{segal:module} and 9 generic (global) regulators obtained
from \citen{lee:binding}.

Due to computational limitations on the number
of (motif,parent) features we could use in training, we select
smaller subsets of
regulators based on the following selection criteria. We use 13
high-variance regulators that had a standard deviation (in
expression over all experiments) above a cutoff of 1.2.
The second
subset consists of the 9 global regulators that are present in
the \citen{lee:binding} studies but absent in the candidate list of \citen{segal:module}.
 We also include 50 regulators that are found to be top
ranking regulators for the clusters identified in Segal et al.
The union of these three lists gives 53 unique regulators.

{\bf\em Target set and label assignment:} We discretize
expression values of all genes into three levels representing
down-regulation (-1), no change (0) and up-regulation (+1) using
cutoffs based on the empirical noise distribution around the
baseline (0) calculated from the three replicate unstimulated (time=0)
heat-shock experiments \citea{gasch:yeast}. We observe that 95\%
of the samples in this distribution had expression values between
+1.3 and -1.3. Thus we use these cutoffs to decide what we
define as significantly up-regulated (+1) and down-regulated (-1)
beyond the levels of biological and
experimental
noise in the microarray
measurements.

\suppress{
It is important to note that we train only on those
(gene,experiment) pairs for which we get a discretization of +1 or
-1, not examples where there is a baseline 0 label.
However, we are able to make predictions on {\em every} example in a
held-out experiment by thresholding on predicted margins, that is,
we abstain from predicting (predict baseline) if a prediction has margin
below threshold (see Results section).
}

Note that although we {\it train} only on those
(gene,experiment) pairs which discretize to up- or down-regulated
states, we {\it test} (make predictions) on every example
in a held-out experiment by thresholding on predicted margins.
That is, we \marginpar{abstaining from predicting is not
the same as predicting baseline-cw}
predict baseline if a prediction has margin
below threshold (see Results section).

We reduce our target gene list to a set of 1411 genes which
include 469 highly variant genes (standard deviation $>$ 1.2 in
expression over
all experiments) and 1250 genes that are part of the 17 clusters
identified by \citen{gasch:yeast} using hierarchical
clustering (eliminating overlaps).

{\bf\em Software:} We use the MLJava software developed by
Freund
and Schapire,
which
implements the ADT learning algorithm. We use the text-feature in
MLJava to take advantage of the sparse motif matrix and use memory
more efficiently.

\section{Results}
\label{sec:experiments}

\subsection{Cross-Validation Experiments}
\label{sec:tenfoldcv}

We first perform cross-validation experiments to evaluate
classification performance on held-out
experiments. We divide
the set of 173 microarray experiments into 10 folds, grouping
replicate experiments together to avoid bias, and perform 10-fold
cross-validation experiments using boosting with ADTs on all 1411
target genes.

We train the ADTs for 400 boosting iterations, during most of
which
test-loss
decreases continuously.
We obtain an accuracy of 88.5\% on
up- and down-regulated
examples averaged over
10 folds (test loss of 11.5\%), showing that
predicting regulatory response is indeed possible in our framework.

To assess the difficulty of the classification task, we also compare
to a baseline method, $k$-nearest neighbor classification (kNN),
where each test example is classified by a vote of its $k$ nearest
neighbors in the training set. For a distance function, we use a
weighted sum of Euclidean distances $d((g_1,e_1),(g_2,e_2))^2 =
w_m ||{\bf m}_{g_1}-{\bf m}_{g_2}||^2+ w_p ||{\bf p}_{e_1}-{\bf
p}_{e_2}||^2$, where ${\bf m}_g$ represents the vector of motif
counts for gene $g$ and ${\bf p}_e$ represents the parent
expression vector in experiment $e$. We try various weight ratios
$10^{-3}<({w_m}/{w_p})<10^3$
and values of $k <
20$, and we use both binary and integer representations of the
motif data. We obtain minimum test-loss of 34.4\% at k=17 for
binary motif counts and 31.3\% at k=19 for integer motif counts,
both for weight-ratios of 1, giving much poorer performance than
boosting with ADTs.

Since ADTs output a real-valued prediction score $F(x) =
\sum_{t=1}^{T}{\alpha_t h_t(x)}$, whose absolute value measures the
confidence of the classification, we can predict
a baseline label by thresholding on this score, that is,
we predict examples to be up- or down-regulated
if $F(x)>a$ or $F(x)<-a$ respectively, and to be baseline if
$|F(x)|<a$, where $a>0$.
\begin{figure}[htb]
\begin{center}
\resizebox{!}{2in}{\includegraphics{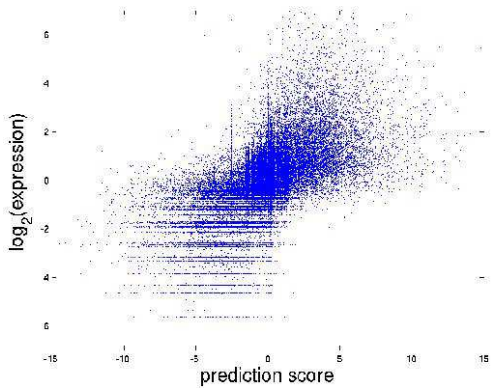}}
\end{center}
\caption{\footnotesize {\bf Scatter plot of true expression values versus prediction scores $F(x)$ .}  The scatter plot shows a high correlation between
prediction scores (x-axis) and true log expression values (y-axis) for genes on held-out experiments.
\label{figure:scatter}}
\end{figure}
Figure \ref{figure:scatter}
shows expression values versus prediction scores for all
examples (up, down, and baseline) from the held-out experiments
using 10-fold cross-validation. The plot shows a
significant
correlation between expression and prediction, reminiscent of the
actual regression task.
(The correlation coefficient is $.74$ for +1 and -1 examples in the
test set and $.59$ for all examples.  While this correlation would not
be considered high for a regression problem, it is significant in our
current setting.)
Assigning thresholds to expression and
prediction values binning the examples into up, down and baseline
we obtain the confusion matrix in Table \ref{table:confusion}.
\begin{table}
\begin{center}
\begin{footnotesize}
\begin{tabular}{ll|ccc}
        &  & \multicolumn{3}{c}{Predicted Bins} \\
    &  & Down & Baseline & Up\\
\hline
           & Down          & 16.5\% & 8.9\% & 1.5\% \\
True Bins  & Baseline      & 9.3\% & 32.4\% & 6.3\% \\
           & Up            & 2.8\% & 9.9\% & 12.0\% \\

\end{tabular}
\end{footnotesize}
\end{center}
\caption{\footnotesize {\bf Confusion Matrix:}
Truth and predictions for all genes in the held-out experiments,
including those expressed at baseline levels. Examples are binned by assigning a threshold $a = \pm0.5$ to
expression and prediction scores.
}
\label{table:confusion}
\end{table}

\subsection{Extracting features for biological interpretation}
\label{sec:biofeatures}
We describe below several approaches for extracting important
features from the learned ADT models, and we suggest ways to relate
these features to the biology of gene regulation.

{\bf\em Extracting significant features:} Features at
nodes in the ADT consist of motif-parent pairs. We rank motifs,
parents and motif-parent pairs by three different methods: by the
boosting iteration in which the feature first occurs ({\em iteration score}),
by the total number
of occurrences of the feature
in the final tree ({\em abundance score}), and by the
absolute prediction score associated to the
feature ({\em prediction score}). Ranking
scores are averaged over all ten folds (see supplementary website
for detailed results). Note that presence of a strong feature does
not necessarily imply a direct binding relationship between parent
and motif.
Such a pair could represent an indirect regulatory relationship
(for example, a kinase and the binding site of the transcription factor
that it phosphorylates) or some other kind of predictive
correlation, for example,
co-occurrence of the true binding site with the motif
corresponding to the feature.

The top ranking motif based on iteration score was the STRE
element of MSN2/MSN4, which is known to be a regulatory element
for a significant number of general stress response
target genes \citea{gasch:yeast}.
The other high scoring motifs include HSF1 (heat-shock), RAP1
(heat-shock and osmolarity), TBP (TATA binding site), ADR1
(glycerol metabolism and osmolarity), MIG1 (glucose metabolism and
carbon source based stress), REB1 (Pol-I transcription termination
activity), GAL4 (galactose metabolism), YAP1 (peroxide stress) and
GCN4 (amino acid biosynthesis and starvation response) binding
sites, all of which are known to be active in various kinds of
stress responses.

\begin{figure}[htb]
\begin{center}
\includegraphics[width=3in]{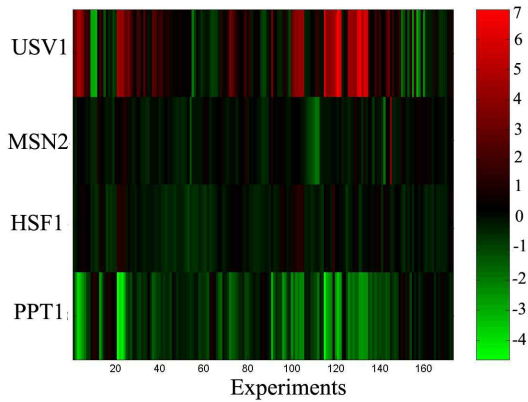}
\caption{\footnotesize {\bf Comparison of expression profiles (173
experiments) of USV1, MSN2, HSF1 and PPT1.}  The mRNA expression
levels of USV1 and PPT1 are informative, with about 50\% and
35\% of experiments (respectively) showing over 2 fold expression
change over wildtype.  The espression levels for MSN2 and HSF1
fall mostly in the baseline state, with only
about 6\% and 5\% of experiments (respectively)
showing at least 2 fold expression
change.  While MSN2 and HSF1 are not identified as high scoring
parents in the learned trees, their binding sites occur as high
scoring motifs.}\label{fig:reg_comp}
\end{center}
\end{figure}

Of the 53 candidate regulators, 37
appear in the ADTs of the ten folds. The top-ranking regulator,
based on both iteration score and abundance score, is USV1
(YPL230W); this regulator was found by \citen{segal:module} as the
top-ranking regulator in 11 of their 50 regulatory modules.
Other top ranking regulators (see Table \ref{table:anshul})
include PPT1, TPK1 (SRA3), XBP1 and GCN20. It is interesting to
note that while the presence and absence of binding sites of some
very important stress factors like MSN2 and HSF1 (heat shock
factor) are identified as significant features (high motif iteration score)
in the ADTs, their mRNA expression levels do not
seem to be very predictive. HSF1 does not appear as a parent in
any of the ADTs, and MSN2 gets low abundance and iteration
scores as a parent, despite its importance as a stress response
regulator. Similar results are observed in the modules of
\citen{segal:module}, where HSF1 is not found in any of the
regulation programs and MSN2 is found in 3 of the 50 regulation programs but with low
significance. If we examine the expression profiles of HSF1, MSN2,
USV1 and PPT1,
we find that
the mRNA levels of MSN2 and HSF1
have quite small fluctuations (rarely greater than 2 fold change) and
fall mostly within the baseline state, while
the expression levels of USV1 or PPT1 show much larger variation over
many experiments (see Figure \ref{fig:reg_comp}).
It is known
that MSN2 is regulated post-translationally by TPK1, which is identified
as an important parent in the ADTs and is found associated with
the MSN2 binding site as a motif-parent pair.
Thus in this case, where the activity of a regulator is
itself regulated
post-transcriptionally, we see a clear
advantage of using motif data along with mRNA expression data.

\begin{table}
\begin{center}
\begin{footnotesize}
\begin{tabular}{c c c c c c}
10 folds&heat-shock&heat-shock &H$_2$O$_2$\\
& & w/o USV1 & \\
\hline
USV1&USV1&SRA3&USV1\\
XBP1&XBP1&XBP1&XBP1\\
SRA3&SRA3&PPT1&SRA3\\
PPT1&PPT1&DAL80&YAP1\\
GIS1&GIS1&GAC1&PPT1\\
YGL099W&SLT2&GIS1&FAR1\\
GAC1&GIP2&SLT2&YGL099W\\
GCN20&GAC1&WTM1&SLT2\\
MTH1&DAL80&SRD1&GAC1\\
HAP4&SRD1&GAT1&MTH1\\
YGL096W&GAT1&GIP2&GIS1\\
\label{table:anshul}
\end{tabular}

\end{footnotesize}
\caption{\footnotesize {\bf Top scoring regulators:} Top scoring
regulators for the 10 fold cross-validation experiment and three
special setups. For additional results on extracted features,
refer to the supplementary website.}
\end{center}
\end{table}

{\bf\em ``In silico'' knock-outs:} By removing a candidate from
the regulator list and retraining the ADT,
 we can evaluate whether test loss significantly
decreases with omission of the parent and identify other weaker
regulators that are also correlated with the labels.  We
investigate in silico knock-outs in the biologically-motivated
experiments described below.

\subsection{Biological Validation Experiments}
\label{sec:biovalidation}
We designed 5 different training and test sets of selected
microarray experiments based on observations of similarity and
differences between stresses by \citen{gasch:yeast}, and we used
these experiments to study the
response to specific types of stress in our framework.
We present results for 3 of
these studies below (see supplementary website for the other 2
experiments). By
comparative analysis of the trees learned from these
sets, we find and validate regulators that are associated to
regulation programs activated by different stresses.

{\bf\em Heat-shock and osmolarity stress response:} In the first
study, we trained on heat-shock, osmolarity, heat-shock knockouts,
over-expression, amino-acid starvation experiments, and we tested
on stationary phase, simultaneous heat-shock and hypo-osmolarity
experiments.

We observe a low test loss of 9.3\%, supporting the hypothesis
that pathways involved in heat-shock and osmolarity stress appear
to be independent
and the joint response to both stresses can be predicted easily
\citea{gasch:yeast}. We also confirm that the regulatory response
for stationary phase (test set) is very similar to that of
heat-shock (training set) \citea{gasch:yeast}. The high scoring
parents are USV1, XBP1, TPK1, PPT1, GIS1, GAC1 and SLT2. The
connection of osmolarity response to the HOG and other 
MAP kinase
pathways is well known, and it is interesting to note that most of
these regulators are in fact signaling molecules. Also, the
osmolarity response is strongly related to glycerol metabolism and
transport
and hence
closely associated with gluconeogenesis and
glucose metabolism pathways. We find the binding sites of CAT8
(gluconeogensis), GAL4 (galactose metabolism), MIG1 (glucose
metabolism), GCN4 (regulator of HOG pathway and amino acid
metabolism), HSF1 (heat-shock factor), CHA4 (amino acid
catabolism), MET31 (methionine biosynthesis), RAP1 and MSN2/MSN4
to have high iteration scores; these regulators are all 
related to the 
stress conditions in the training set.

{\bf\em USV1 ``in silico'' knockout:} \label{sec:insilico}
Using the same train and test microarrays as in the
heat-shock/osmolarity setup, we perform a second study by removing
one of the strong regulators, USV1, from the parent set and
retraining the ADT.
We get a minor but significant increase in test error from 9.3\%
to 11\%. Regarding structural changes in the ADT, we observe that
the overall hierarchy of the features does not change
significantly: TPK1, XBP1, PPT1 and GIS1 remain the highest
scoring parents. We also find that 305 target genes change
prediction labels. GO annotation enrichment analysis of these
target genes reveal the terms cell wall organization and
biogenesis, heat-shock protein activity, galactose, acetyl-CoA and
chitin metabolism and tRNA processing and cell-growth. These match
many of the terms enriched by analyzing GO annotations of genes
that changed significantly in a microarray experiment by
\citen{segal:module} with stationary phase induced in a USV1
knockout.

{\bf\em Pleiotropic response to diamide:} For the third study, we
trained on heat-shock, heat-shock knockouts, over-expression,
H$_2$O$_2$ wild-type and mutant, menadione, DTT experiments, and
we tested on diamide experiments.

\citen{gasch:yeast} consider the diamide
response to be a composite of responses to the experiments in the
training set. We observe a moderate test loss of 16\%, suggesting
that this pleiotropic response is more complex than the simpler
additive responses to heat-shock and osmolarity. We observe the
emergence of an important motif-parent feature: YAP1 is directly
associated with its ARE-binding site (Y\$TRX2) as a high scoring
regulator that is absent in the ADTs of both previous studies.
This finding is consistent with known biology, since YAP1 plays a
specific role in peroxide and superoxide response (present in the
training set)~\citea{gasch:yeast}. We also find the PDR3
(menadione-drug response) binding site to be a high scoring motif.

\section{Discussion}
\suppress{
\begin{enumerate}
\item Discuss thresholding issues: how to deal correctly with genes that are only baseline or up

\item Extend MLJAVA to deal with real-valued sparse features?

\item Other issues?
\end{enumerate}
}
While encouraged by the performance of our method, we believe
further work is likely to yield much more comprehensive and accurate
models of the regulatory networks of yeast and other
simple organisms.

One main direction for improvement is to increase the
computational efficiency of our software so that we can scale up
both the size of the training set and the feature space.
Since the
Gasch dataset that we used here only contains experiments
for
environmental stress response, 
many other
regulatory pathways are likely not activated and therefore cannot be modeled
by analysis of this dataset alone.
We plan to
pursue more extensive computational experiments on other diverse
yeast datasets, such as those available through NCBI's Gene
Expression Omnibus and the Saccharomyces Genome Database (SGD). At
the same time, we
hope to
increase the number of parents
to include the complete putative list of about 500 regulators,
in order to
identify the possible roles of
additional regulatory proteins.
Since we are using (motif, parent) feature pairs, increasing the
number of parents increases the feature space and memory requirements
by a multiplicative factor.
Two promising directions for improvement are (i) using data structures
more appropriate for our pairwise interaction features and (ii) using
weighted sampling to reduce the size of the memory required for
the training data.

Another potential advance would be a more careful treatment of the
raw data.  In these preliminary experiments, we used
a simple noise model based on wildtype replicates, and we were able
to learn to predict large up- or down-regulation response using thresholds
based on this model.
However, while log expression ratio data (perturbation/wild type) gives a
natural input variable for our analysis, better signal to noise
is likely to be achieved by taking into account the excitation levels
separately.  In particular, using an intensity-sensitive noise model
could allow us to establish more meaningful thresholds for more of the
genes.  A more complicated issue is the fact that
we do not use baseline expression
examples for training, and therefore we restrict to the subset of
genes that show variation across stress response experiments for our
training and test sets.  Ideally, however, we would like to predict
regulatory response for all genes (including non-responding genes), which
will likely mean changing the formulation of the learning task so that
we include baseline examples in training.

A further refinement
would be to treat parent and child
excitation levels as continuous rather than binary quantities.
Similarly, the number of motifs
in the regulatory region,
rather than merely their presence/absence,
and the spatial relationship between them
could be taken into account. While these
extensions
could potentially yield much
more realistic models, they require substantial
algorithmic changes
and should be done carefully so as to avoid overfitting.

While we showed how to extract significant motifs, regulators, and
motif-regulator pairs from the ADTs, further work is needed
to obtain more detailed information from these
predictive models.
It is plausible that the trees contain information about combinatorial
relationships between regulators or between regulatory elements, but it
is not clear how to disambiguate independent effects from combinatorial
ones. One possible approach could be to rank collections of
two or more features occurring in paths in the ADTs and check whether
motifs in over-represented paths tend to co-occur in regulatory regions,
giving evidence of combinatorial relationships.  Another approach is
to examine more carefully
the contributions of different features to the prediction
score for various target genes.  While our learning method does not yield
a descriptive network model that can be easily visualized, we believe
that the predictive model approach enables new possibilities for analysis
and understanding of gene regulation.

\suppress{
\item  Other large-margin classifiers: the success evidenced by alternating decision trees begs the question: how do other modern machine learning margin-based classifiers perform within this framework? In particular, we are working to develop a support vector machine approach
\item  Brown Boosting (Yoav)
}

\section{Acknowledgement}
AK is supported by NSF EEC-00-88001.
CW and MM are partially supported by NSF
ECS-0332479 and NIH GM36277.
CL and CW are supported
by NIH grant LM07276-02, and
CL is supported by an Award in Informatics
from the PhRMA Foundation.

\bibliographystyle{harvard}

\suppress{

}
\end{document}